# Nanophotonic waveguide enhanced Raman spectroscopy of biological submonolayers.

Ashim Dhakal[*,1, 2, 3], Pieter C. Wuytens[1,2,4], Frédéric Peyskens[1,2], Karolien Jans[3], Nicolas Le Thomas[1,2] and Roel Baets[1,2]

[1]Photonics Research Group, INTEC Department-Ghent University-imec, 9000 Ghent, Belgium,

[2]Center for Nano- and Biophotonics, Ghent University, 9000 Ghent, Belgium

[3]imec, Kapeldreef 75, 3001 Heverlee, Belgium

[4]Department of Molecular Biotechnology, Ghent University, 9000 Ghent Belgium

KEYWORDS. Raman spectroscopy, Integrated Photonics, Channel Waveguides, Photonic sensors, Slotted wavguides

**ABSTRACT:** Characterizing a monolayer of biological molecules has been a major challenge. We demonstrate nanophotonic waveguide enhanced Raman spectroscopy (NWERS) of monolayers in the near-infrared region, enabling real-time measurements of the hybridization of DNA strands and the density of sub-monolayers of biotin-streptavidin complex immobilized on top of a photonics chip. NWERS is based on enhanced evanescent excitation and collection of spontaneous Raman scattering near nanophotonic waveguides, which for a one centimeter silicon nitride waveguide delivers a signal that is more than four orders of magnitude higher in comparison to a confocal Raman microscope. The reduced acquisition time and specificity of the signal allows for a quantitative and real-time characterization of surface species, hitherto not possible using Raman spectroscopy. NWERS provides a direct analytic tool for monolayer research and also opens a route to compact microscope-less lab-on-a-chip devices with integrated sources, spectrometers and detectors fabricated using a mass-producible CMOS technology platform.

Monolayers or thin layers of materials are ubiquitous in nature and play a critical role in modern instrumentation and technology[1-3]. The study of monolayers at the bio-interfaces is crucial for applications ranging from clinical diagnostics, genomics, proteomics, and biomaterials to tissue engineering [1-8]. Furthermore, there is an emergence of numerous interesting physical processes and novel optoelectronic applications that use 2D monolayers of materials such as graphene[9] and graphene-like 2D materials[10-11]. Hence, a paramount importance is placed on understanding the structural, physical and biochemical properties of monolayers or very thin layers of molecules.

In monolayer research, there is a need for a precise tool that provides consolidated information about composition, molecular structure, and density of the surface molecules. Simultaneously characterizing the kinetics of the bio-chemical interactions at interfaces is equally important in many applications[1-8], such as the study of DNA hybridization. Obtaining such consolidated information from monolayers is particularly challenging because only a small number of molecules contribute to the signal. The prevalent techniques such as scanning probe microscopy (SPM)[12], X-ray based techniques[13-15] and nuclear magnetic resonance microscopy[16] are often invasive, expensive, necessitate a long time for sample preparation and data acquisition, and require a restrictive environment for samples. Some techniques such as surface plasmon resonance (SPR)[17] and whispering gallery mode (WGM)[18] sensors measure adsorption reactions via resonance shifts due to minute changes in the refractive index near a surface. These techniques are prone to errors occurring from non-specific binding because the signals lack direct molecular signature. As such, many of these existing techniques for monolayer characterization are unsuitable for a variety of applications.

Raman spectroscopy is a powerful technique, and is particularly well-suited for identification and quantification of various physicochemical properties of molecules. Raman spectroscopy has been successfully used for the characterization of monolayers of materials such as graphene that have a high Raman scattering efficiency[19] or via a pump wavelength resonant to the fluorescence that allows for a huge enhancement in the Raman signal[20]. Very thin layers of small Raman cross-section molecules, such as nucleic acids, proteins or lipids, present a major challenge for this technique, because the associated signal acquisition times to obtain a sufficient signal-to-noise ratio (SNR) are impractically long. This is especially case for near-infrared (NIR) wavelengths, where autofluorescence background from most of the biological materials is weak, yet, unfortunately the Raman scattering cross-sections are also very small. Surface enhanced Raman scattering (SERS) has been used to mitigate this problem[21-22] by means of electromagnetic enhancement due to surface plasmon resonance effects. SERS signals depend strongly on geometrical features at the nanoscale and the resultant variations from one substrate to another necessitate sophisticated fabrication techniques for reproducibility and uniformity[23-25]. Tip enhanced Raman spectroscopy (TERS), a technique combining SPM with a SERS nano-tip to greatly enhance the signal, has been recently used to study polymer monolayers[26], albeit with the disadvantages of long integration time and inhomogeneity of signal strength inherent to SPM and SERS.

The potential of waveguiding to enhance Raman signals has been recognized[27] as early as 1980. A planar waveguide was utilized[28] to evanescently excite the monolayers adsorbed on the top-cladding, with a microscope needed to collect the Raman signal from the top of the waveguide. Enhancement was only moderate; hence, implementation of resonance Raman scattering was necessary to obtain a reasonable signal-to-noise ratio. These techniques, along with all other techniques discussed so far, are not easily integrable on a chip to allow for mass fabrication of a low cost and microscope-less lab-on-a-chip device. Recently, we reported a proof-of-concept lab-on-a-chip approach for quantitative spontaneous Raman sensing of bulk liquids[29-31]. This technique exploited electromagnetic interaction of the molecules with the evanescent tail of the fundamental mode of nanophotonic waveguides[29-33]. Here we report a real-time observation of immobilized DNA hybridization and an accurate and direct measurement of the density of sub-monolayers of biotin-streptavidin complex using this on-chip nanophotonic waveguide enhanced Raman spectroscopy (NWERS) technique. This prospect emerges as a consequence of electromagnetic enhancement and an increase in number of the probed molecules, bringing about an enormous reduction of the integration time for a practical SNR. The following section describes this matter in detail.

**Enhanced evanescent Raman interaction**

A schematic of the basic device is shown in Fig. 1(a). The fundamental TE mode of a waveguide with an excitation wavelength $\lambda_o$ (785 nm in this article) evanescently excites the layer of the molecules at the waveguide core-cladding interface. The spontaneous Raman scattered light emitted by the molecules is collected via the same waveguide. For a molecule at a given position $r_o$, the strength of excitation and collection depends on the fourth power of the power-normalized modal field strength at that position[30]. Due to the electromagnetic confinement, the power $P_{w,o}$ collected from a molecule at a position very close to the surface of the waveguide will be relatively large compared to the free-space excitation and emission, especially in high contrast waveguides. To compare with ideal free-space or microscopic techniques, the power of the Raman signal collected by the waveguides can be normalized by the total Raman power emitted by the same molecule when excited by the average intensity of an ideal diffraction limited beam (NA = 1) carrying the same pump power in free space (see supplementary information (SI) section 1 for details). The normalized collected power $\bar{P}_{w,o}$ is shown in Fig. 1(b)-(c) for the fundamental TE mode of silicon nitride ($Si_3N_4$) waveguides[34] excited at $\lambda_o$ and emitting sufficiently close (< 100 nm, see SI) to $\lambda_o$. The figures indicate that, depending on the geometry of the waveguides, the regions near to the waveguides have comparable and even greater power conversion efficiency than the most ideal free space excitation and collection in a microscope. The fact that $\bar{P}_{w,o} > 1$ is indicative of a broadband Purcell enhancement present in the waveguides[35]. In the SI section 1, we observe that $\bar{P}_{w,o} > 10$ for silicon nitride slot waveguide with slot width $s = 20$ nm indicating that a substantial signal enhancement is possible with those waveguides.

In addition to the transverse enhancement, the most significant component of the enhancement of the waveguide approach comes from the longitudinal propagation of the waveguide mode (Fig. 1(d)). The effective interaction area contributing to the Raman signal in a guided mode $A_{eff,w}$ is almost equal to the total surface area $A_{wg}$ of the waveguides which can be made arbitrarily large by using longer waveguides, and are limited only by waveguide losses. With the current technology, waveguide losses of around 1 dB cm$^{-1}$ are typical[30] thereby allowing for waveguide lengths in the order of several centimeters without a significant loss of the pump or the Raman signal. In contrast, for a diffraction-limited beam, the effective probed surface is determined by the waist of the focused beam $w_o$ (Fig. 1(e)).

For the case of bulk materials placed on the top of the waveguides, an immediate consequence of these results is that the NWERS approach provides a signal more than two orders of magnitudes higher than via the usual microscopic methods[29-31, 33]. The advantage of NWERS compared to the free-space approach becomes more prominent for a monolayer of molecules functionalized on top of the waveguides. The enhanced evanescent field and the extended area of interaction along the waveguide length $l$ lead to a very high Raman signal from the monolayers adsorbed on the waveguides. The power collected by the waveguides ($P_{w,s}$), normalized to the total emitted power ($P_{g,s}$) for free space diffraction-limited excitation for the same surface density $\rho_s$ and scattering cross-section $\sigma$ of molecules is given by:

$$\bar{P}_{w,s} = \frac{P_{w,s}}{P_{g,s}} = l\eta_s \qquad (1)$$

where, $\eta_s$ is the *surface conversion efficiency* which is a function of the distribution of the modal field at the waveguide surface (see SI section 1). It can be calculated using standard mode solvers and can be tuned by designing an appropriate waveguide geometry. Fig. 2(a) provides the calculated $\eta_s$ as a function of $Si_3N_4$ waveguide width $w$ for several slot widths $s$ for waveguides (see Fig 1(a) for the schematic of the waveguide cross-section). Depending on the design of the waveguides, Fig 2(c) illustrates that for a 1 cm $Si_3N_4$ waveguide we can theoretically expect a signal that is 4 to 5 orders of magnitude higher compared to the free space case. The enhancement can be further improved by using higher index-contrast waveguides and more efficient waveguide designs[30].

Fig. 1(b)-(c) also shows that the collected signal decays exponentially as a function of the distance away from the surface of the waveguide. For the $Si_3N_4$ waveguides with water as the top cladding, the total signal contribution halves almost every 20 nm away from the waveguides. A small depth of field is a valuable characteristic of the NWERS approach, as it results in a well-defined excitation and detection volume next to the surface of the waveguide.

**Experimental demonstration of waveguide enhancement**

To measure the Raman signal from waveguides and compare with a commercial Raman microscope, we functionalize a monolayer of Rhodamine molecules on an amino-silanized $Si_3N_4$ waveguide surface (see *Methods*). The setup used to measure the collected Raman signal from the waveguides has been described elsewhere[29, 25] in detail. A pump with 30 mW power is coupled to the waveguide with 8±2 dB coupling loss per facet. In the following part of this article, unless stated otherwise, we use a waveguide width $w$ = 850 nm slotted waveguide with slot width $s$ = 150 nm (Fig. 1(c)). We prefer to use slotted waveguides over striped ($s$ = 0) waveguides because our slotted waveguides have comparable losses (~1.3 dB/cm compared to ~ 0.5 dB/cm in water), more than 3 times higher conversion efficiency, and lower background[49]. The co-propagating light is collected, filtered and coupled to a single mode fiber to measure the Raman spectrum with a commercial spectrometer (*Andor* SR303i) and a cooled CCD detector (*Andor iDUS 416*). The measured spectrum is displayed in Fig 2(b). For NWERS measurements, an integration time of 4 s is sufficient to obtain SNR of 50 or more. The noise is mainly dominated by the shot noise of the Raman signal from the analyte and from the background associated with the waveguide material.

Figure 2(b) also shows the spectra measured from a Rhodamine-functionalized Raman-grade calcium fluoride ($CaF_2$) slide measured using a commercial Raman confocal microscope (see *Methods*) and a CCD detector with similar characteristics. To be able to measure much weaker Raman signals from the monolayers using a microscope, it is important to have a low background material because the microscope cannot discriminate the background from the other materials in the confocal volume. We use Raman grade ($CaF_2$) microscope slides for the measurements with the Raman microscope as it exhibits very low Raman background compared to the stack of materials used for the fabrication of the waveguides (see *Methods* for fabrication details). A 7±2 nm layer of similar $Si_3N_4$ is deposited on $CaF_2$ slides and functionalized together with the $Si_3N_4$ waveguide samples to ensure that both of the samples have similar density of molecules. For the spectra obtained with the Raman microscope, the noise is mainly dominated by the dark noise of the detector, as the signal is very low. A pump power of 18 mW and integration time of 80 s was required to obtain a SNR of around 8.

Thus, even ignoring a total of 18 dB coupling losses (in this specific case), we clearly observe a very strong enhancement of the Raman signal from the waveguides. We normalize the observed signal $P_s$ with the transmitted pump power $P_{tx}$ to correct for the variations in the coupling losses in the system which are extrinsic to the NWERS approach and can be eliminated by optimized coupling mechanism or by integrating the sources and detectors on the chip. The normalized collected power $\zeta$ is directly related to the intrinsic variables of the system[27] such as the difference in the waveguide losses for pump and Stokes wavelengths $\delta\alpha$, the length of the waveguide $l$ and surface efficiency $\eta_s$ :

$$\zeta \equiv \frac{P_s}{P_{tx}} = \frac{\eta_s}{2}\sigma\rho_s l(1+\Delta) \qquad (2)$$

Here, $\Delta = (l\,\delta\alpha)/2! + (l\,\delta\alpha)^2/3! +..$ is generally negligible for small waveguide lengths, small waveguide losses or small stokes shifts. In a later section we show how the effect of $\Delta$ can be determined for an accurate measurement. Figure 2 (c) displays the SNR versus $\zeta$ graph for the Raman signal calculated for the 1355 cm$^{-1}$ line of monolayers of Rhodamine using the data presented in Fig. 2(b). The values are normalized for the 18 mW pump power and 80 s integration times after the correction for the coupling losses in order to provide a comparison with the microscope system. The estimated values have been repeatable for at least three different samples with less than 25% variations.

Figure 2(c) confirms more than four orders of magnitude enhancement in the signal from only one centimeter of waveguide length, in accordance with the expectation from the theoretical curve shown in Fig. 2(a). In addition, we see that compared to commercial microscopic systems, the enhancement of the NWERS signal leads to more than two orders of magnitude improvement in SNR. Figure 2(d) illustrates the evolution of SNR as the density of the molecules or the cross-section of the molecules is varied, also calculated using the data shown in Fig. 2(b) (see SI Sec 2). More than two orders of magnitude improvement of SNR or limit of detection (LoD) is expected with the NWERS for the same pump power and same integration time. If the shot noise from the waveguide background can be reduced, the improvement of SNR or LoD can approach to four orders of magnitude. Seen from a slightly different perspective, depending on the dominating source of noise and the concentration of analyte, the NWERS approach leads to 4 to 8 orders (see SI Sec 2) of reduction in the integration time compared to the free-space system for a similar SNR and pump power. Further, as can be seen in Fig 2(a), the surface efficiency value $\eta_s$ is practically invariant with small variations in the waveguide dimensions that might occur during fabrication (<4% for a variation of 20 nm); hence $\eta_s$ will remain very close to the calculated value in average[30]. Our results for NWERS break the impasse of impractically long integration times and unpredictable signal enhancement that several potential applications of Raman spectroscopy are facing. Such an asset of NWERS is implemented for two different applications described subsequently in this paper. In the next section, we outline real-time observation of DNA hybridization on the surface of silicon nitride waveguides using NWERS.

**DNA hybridization kinetics using spontaneous Raman signal**

DNA microarrays are indispensable tools in modern biotechnology with a broad range of applications from gene expression profiling and drug discovery to forensics[5-6]. A reliable, cost-effective and sensitive quantification method for real-time analysis of DNA hybridization has been identified as a key necessity to broaden the range of applications of the DNA microarrays[36,6]. Here we demonstrate a real-time analysis of DNA hybridization using NWERS, with an enormous prospective for large scale integration and parallelization.

Oligonucleotide strands (DNA: 5'- /hexynyl/-TTT TTT TTT TCA CCA GCT CCA ACT ACC AC -3') of K-Ras gene - an important gene the activation of which is responsible for 17-25% of all human cancer tumors[37], are immobilized on our chips using *copper-catalyzed alkyne-azide 1,3-dipolar cycloaddition* reaction on a silanized surface (see *Methods*). The density of DNA probes was estimated to be about $6\pm1 \cdot 10^{12}$ cm$^{-2}$, based on measurement of P-concentration by total X-ray fluorescence spectroscopy. A 500 nM solution of cDNA with cy3 marker (5'- /cy3/- GTG GTA GTT GGA GCT GAA AAA AAA AA -3'), in a 0.5M NaCl/TE buffer is used as analyte for the hybridization process.

Initially, a 150 μl of buffer is drop-casted on the chip to measure the Raman spectra from the chip with immobilized hexynyl-DNA in buffer. At time $t$ = 60 s, 15 μl of cDNA is added on the chip and the changes of the Raman spectrum are monitored until a stationary regime is reached. The changes in the Raman spectrum are directly related to the hybridization kinetics (see SI section 3 for details) hence a stationary Raman spectrum indicates a saturation of the reaction. The hybridization reaction saturates in about 250 s after application of the cDNA solution. Figure 3(a) shows the spectrogram of the observed Raman spectra collected from the chip as it evolves during the hybridization process. Each spectrum is measured every two seconds with about 0.25 mW effective power in the waveguide and corrected for the background from the waveguide (SI section 5). Fig. 3(b) shows the Raman spectra of the immobilized hexynyl-DNA complex in buffer (at $t$ = 6 s) and Raman spectra of the DNA hybridization duplex (hyxynyl-DNA·cDNA-Cy3) after the hybridization reaction saturates ($t$ = 250 s). The reaction kinetics can be quantified via one of the peaks corresponding to cDNA. In our case, we choose the 1392 cm$^{-1}$ line of cy3 for its distinctive nature.

As seen in Figure 3(c), the reaction kinetic observed using the 1392 cm$^{-1}$ Raman line fits well with the first order Langmuir equation[38-39] (SI section 3): $\zeta(\tau) = \zeta_\infty \{1-\exp(-K_1\rho_A\tau)\}$. In the present case, $\zeta(\tau)$ is the normalized Raman signal intensity collected through the waveguide at a time $\tau$ after the start of the reaction, $\zeta_\infty$ is the normalized Raman signal corresponding to the total number of binding sites i.e. Raman signal at the time of saturation, $\rho_A$ is the concentration of the reacting analyte on the chip ($\rho_A$ = 45.4 nM in this study). The least-square-error fit to the above equation yielded a goodness of fit $R^2$ = 0.95. The association constant $K_1$ is estimated to be $2.9 \pm 0.6 \cdot 10^5$ M$^{-1}$ s$^{-1}$ from the fit. A higher temperature ~ 26° C of our experiments may explain the slightly higher value we obtained compared to values reported in the literature (1.2·10$^5$ M$^{-1}$ s$^{-1}$ for 20° C) [38-39] under otherwise similar experimental conditions such as chain length, target concentration and probe density.

Although cDNA with a marker label is used in this proof of concept experiment, a natural next step is a label-free measurement of hybridization kinetics using NWERS. For this purpose changes in the Raman spectra of the nucleic acids during hybridization have to be identified and quantified. The current spectra are overwhelmed by the Raman signal from cy3 dye, hence are not suitable for identification of such changes during hybridization.

There are several indirect techniques for the study of hybridization kinetics[38-40]. The appeal of the proposed NWERS is its simplicity and a direct correspondence of the spontaneous Raman signals with the number of the probed molecules. This makes NWERS intrinsically robust against the non-specific binding, photo damage, and photo bleaching that constitute a major source of error in most of the existing techniques, such as those based on fluorescence, X-rays or a microbalance.

Since $\zeta_\infty$ is the normalized Raman signal corresponding to a known probe density, we can use Eq. (2) to accurately determine the spontaneous scattering cross-section of the analyte molecules corresponding to a Raman peak (see SI section 4). Once the cross-section of a molecule is determined, the NWERS technique can be used to calculate the loading density of the probes, in a future experiment, without a need for determination of probe density that would otherwise require specialized techniques such as XFS. In the next section we discuss this application of NWERS for a robust quantification of the surface loading in another important biological assay based on biotin-avidin binding.

**Quantification of surface loading for a sub-monolayer of biotin and biotin-streptavidin complex**

Thanks to its unique structure and small size, biotin shows specific and very strong non-covalent binding to particular proteins such as avidin and histone, without significantly altering the biological activity of the molecules in its surrounding[41-44]. Hence, biotin-based assays have been an indispensable tool in modern biotechnology[41] for the detection and localization of specific proteins, nucleic acids, lipids, and carbohydrates. Here, we demonstrate that NWERS can allow direct detection of amino-silane precursor generally used for functionalization, biotin, and NeutrAvidin and can be used for quantification of the biotin-StreptAvidin loading density when applied to the biotin based assays.

To this end, we incubate biotinylated chips (see *Methods*) with unconjugated NeutrAvidin and StreptAvidin conjugated with Rhodamine red. These avidin derivatives are known to bind with high specificity to the biotinylated surface of the chips[43]. Figure 4(a) depicts a schematic of molecules attached to the waveguides at various stages of functionalization process. Typical Raman spectra obtained from the chips corresponding to these stages are shown in Fig 4(b). We see several peaks corresponding to the molecules that are present at the specific stages of functionalization. Specifically, spontaneous Raman signals from monolayers of functionalized 3-*aminopropyltriethoxysilane*[45] – (henceforth, referred to as amino-silane (AS)), biotin[22,46] covalently bound to AS (AS-B), NeutrAvidin[46] non-covalently bound to functionalized biotin (AS-B·NA), and StreptAvidin-Rhodamine[43-44] conjugate bound to functionalized biotin (AS-B·SA-Rh) can all be detected with acquisition times in the order of a few seconds with a SNR >10.

Functionalized AS display broadband features centered at ~ 1600 cm$^{-1}$, ~ 1415 cm$^{-1}$ and ~1165 cm$^{-1}$ respectively assigned[45] to amide deformations, C-N stretch in primary amines, and $CH_2$ deformations. The important features[22,46] of the AS-B spectra are centered around ~1310 cm$^{-1}$ ($\gamma$-$CH_2$), ~1442 cm$^{-1}$ ($\delta$-$CH_2$, $\delta$-$CH_3$) and ~1630 cm$^{-1}$ (Biotin Ureido ring stretching). Similarly, the spectra with biotin-avidin[22,46] contained all the features of the biotin precursor and some extra peaks centered at ~1132 cm$^{-1}$ (Trp W7), ~1250 cm$^{-1}$ (Amide III), ~1550 cm$^{-1}$ (Trp W3), ~1668 cm$^{-1}$ (Amide I). The spectrum due to AS-B·SA-Rh complex is overwhelmed by the spectral features of Rhodamine, which has more than an order of magnitude higher cross-section compared to that of the precursor molecules. Incidentally we note that, under assumptions of identical loading density across all the functionalized chips, the respective peak amplitudes assigned to different molecules can be used to estimate their relative Raman cross-sections.

To accurately determine the number of streptavidin molecules attached to the surface of our chips, we measure the signal from streptavidin–Rhodamine red conjugate from the biotinylated chip containing the waveguides of lengths 1 cm, 2 cm 3 cm and 4 cm. The normalized collected power $\zeta$ (for the 1513 cm$^{-1}$ Rhodamine line) was measured for different lengths of waveguides using a spectrometer with calibrated CCD. The chip containing Rhodamine is selected, as Rhodamine is a well-studied Raman reporter and its cross-section is well documented. From literature we estimate $\sigma$ for the 1513 cm$^{-1}$ line of Rhodamine used for the experiment to be about 2.8·10$^{-27}$cm$^2$·sr$^{-1}$·molecule$^{-1}$ at 785 nm[47-48] and use $\eta_s$ = 2.1·10$^4$ sr·cm$^{-1}$. Using these values of $\sigma\eta_s$ and the measured $\zeta$, we can determine experimental values of $\rho_s (1+ \Delta) = 2\zeta/(l\sigma\eta_s)$ for different waveguide lengths. These values are nearly equal to the number of molecules adsorbed on the waveguides, aside from the contribution of the factor $\Delta$ shown in Eq. (2). To determine this contribution we fit the observed data with Eq. (2) as a model using the least-square error fitting algorithm. Fig 4(c) shows the plot for $2\zeta/(l\sigma\eta_s)$ as a function of waveguide length $l$ with respective experimental errors. The dominating source of error in our measurements is the vibration of the coupling mechanism[30]. A goodness of fit $R^2$ = 0.995 is obtained and yields a surface loading of 1.0 ± 0.2·10$^{11}$ molecules/cm$^2$ and $\delta\alpha$ = -1.3 ±0.2 dB/cm. Our value for surface density is about 1% of a closely packed monolayer of streptavidin molecules (~1·10$^{13}$ molecules/cm$^2$)[49]. Thus, we see that neglecting $\Delta$ leads to about 15% underestimation of surface density values for a 1 cm waveguide for ~105 nm difference of the pump and Stokes wavelengths, and for waveguide loss of ~2.5 dB/cm at 785 nm.

As a proof-of-concept demonstration, here we focused on the biologically relevant problem of detection and quantification of biotin and DNA assays. We emphasize that there are several surface immobilization techniques developed for silica and silicon nitride surfaces[50] which can be readily employed for the analysis of different monolayers using NWERS.

## DISCUSSION AND CONCLUSIONS

The NWERS-approach allows us to monitor biochemical reactions in real time and to extract consolidated quantitative information about the surface species, such as their chemical composition, molecular structure, loading density or their Raman cross-sections. Because these measurements are based on a highly specific Raman signal from the analyte molecules themselves, assuming that the analyte spectrum is distinct compared to background molecules, our technique is inherently robust against any non-specific binding that may occur during the binding process, unlike most of the existing non-Raman techniques. Furthermore, spontaneous Raman techniques are robust against problems like photo-bleaching, photo-damage, and sample heating, and do not need a restrictive sample environment such as a vacuum for operation.

As a result of the enhancement, mostly originating from the longitudinal propagation along the waveguide, the NWERS system out-performs the commercial microscopic systems in terms of SNR or LoD for a given pump power and integration time. Due to the exponential nature of the evanescent wave, the background light originating from any irrelevant volume also poses little problem. However, the detection limit for a very low concentration can be improved further if the existing background from the waveguides itself can be reduced. The deposition method of the waveguide $Si_3N_4$ has been improved significantly to reduce the background, and can possibly be improved further. The use of alternative waveguide designs such as the slotted waveguides with even narrower slots may further reduce the background compared to the signal diminishing the LoD further.

From the data shown Fig 2(b), the estimated Raman efficiency ($\sigma_s\rho_s$) 1-sigma LoD for the current system is about $2 \cdot 10^{-19}$ sr$^{-1}$.

In conclusion, we have demonstrated the use of integrated single-mode waveguides for evanescent excitation and collection of spontaneous Raman scattering from sub-monolayers. The method leads to an enhancement of at least four orders of magnitude of the spontaneous Raman signal for just a centimeter of waveguide, relative to a standard confocal Raman microscope. This allows for a reduction of the integration times to the sub-second in the NIR region with a reasonable SNR and thereby opening up a path for real-time analysis of biological interactions at interfaces via spontaneous Raman signals. Furthermore, the smallest possible étendue of the pump and collected light due to the use of single-mode waveguide ensures efficient integration with the most compact photonic components such as lasers, integrated spectrometers, filters, and detectors thereby potentially eliminating all the bulk optics and the associated insertion loss, cost, complexity, fragility, volume and weight. The accumulated benefits of design flexibility, simple integration, high performance, possibility of mass-fabrication, compactness and immunity from unwanted electromagnetic interference are all indicative that the NWERS can trigger a plethora of novel applications, including point-of-need Raman analysis.

## METHODS

**Fabrication of the waveguides.** $Si_3N_4$ waveguide circuits used for the experiments described in this article are fabricated on a 200 mm silicon wafer containing a stack of 2.2 μm -2.4 μm thick high-density plasma chemical vapor deposition silicon oxide ($SiO_2$) and 220 nm thick plasma-enhanced-CVD $Si_3N_4$. The structures were patterned with 193 nm optical lithography and subsequently etched by the fluorine based inductive coupled plasma-reactive ion-etch process to attain the final structure[28].

**Silanization.** Formation of covalently bonded amine group on the $Si_3N_4$ waveguides was carried out using well-established 3-aminopropyltriethoxysilane (APTES) based amino-silanization chemistry[50-51]. Samples were cleaned in acetone and isopropyl alcohol, and oxidized in Piranha solution ($H_2SO_4:H_2O_2$, 7:3) at 50°C for 1 hour, to expose hydroxyl groups on the surface. The samples were then incubated in 1% APTES solution in dry toluene for 4 hours in cleanroom (CR) conditions, sonicated for 5 minutes, rinsed thoroughly in dry toluene and deionized (DI) water, and finally cured at 100° C in vacuum for one hour.

**DNA immobilization.** The azide-silane was deposited by vapor phase deposition in a Thermo-Scientific vacuum oven. Silanization occurred at 140 °C and 25 mbar. The azide-SAM modified samples were incubated for 1h in a humidity chamber with a solution of 33.3 % of 50 μM hexynyl-DNA in DIW, 22.2 % of 2 mM *Tris[(1-benzyl-1H-1,2,3-triazol-4-yl)methyl]amine* (TBTA) in *dimethyl sulfoxide* (DMSO), 22.2 % of 2 mM *tetrakis(acetonitrile) copper(I) hexafluorophosphate* (TCH) in DMSO and 22.2 % of 2.6 mM *Sodium L-ascorbate* SA in DI water. After incubation, the samples were rinsed toughly with DMSO. DNA hybridization was performed by incubating the samples in a 500 nM solution of Cy3-labeled complementary DNA strands in hybridization buffer.

**Rhodamine immobilization and biotinylization.** Standard *N-Hydroxysuccinimide* (NHS) ester based chemistry[40] uses NHS ester-activated compounds that react with primary amines on the silanized samples in physiologic conditions to yield specific and stable bonds. The samples were immersed in the 0.1mg/ml NHS-Rhodamine or NHS-Biotin solution in phosphate buffered saline (PBS) at pH 7.2 for 4 hours at CR conditions and rinsed thoroughly with PBS and DI water. The highly specific chemistry ensures fixation of a monolayer of the Rhodamine and biotin molecules on the silanized samples via amide bonds.

**Biotin specific immobilization.** NeutrAvidin and Streptavidin-Rhodamine complex are fixed onto biotinylated surface of the chips by immersing the samples in the 0.1mg/ml solution in PBS for 4 hours at CR conditions and rinsed thoroughly with PBS and DI water.

**Raman microscope.** A WITec Alpha300R+ confocal Raman microscope equipped with a Zeiss W Plan-Apochromat VIS-IR 63x/1.0 objective, a 785 nm excitation diode laser (Toptica) and an UHTS 300 spectrometer using a -75 °C cooled CCD camera (ANDOR iDus 401) was used. A fiber with 100 μm diameter was used as a pinhole.

**Materials.** NHS-Biotin, NHS-Rhodamine, NeutrAvidin and Streptavidin-Rhodamine conjugates were purchased from ThermoFisher-scientific. Labelled cDNA was purchased from Integrated DNA Technologies (IDT). Unless stated otherwise, all other materials were purchased from Sigma Aldrich, and all materials were stored and used as recommended by the manufacturer.

## ASSOCIATED CONTENT

**Supporting Information**. This material is available free of charge via the Internet at http://pubs.acs.org.".

## AUTHOR INFORMATION

**Corresponding Author**

* ashim.dhakal@ugent.be

**Author Contributions**

RB and NLT directed the research. RB and AD conceived the idea. AD developed the theory, designed the experiment and the waveguides, built the experimental set-up, optimized the bio-functionalization process, performed the experiments and wrote the original manuscript. PW carried out the measurements involving Raman microscope. FP calibrated the spectrometer. KJ optimized the DNA immobilization process. All authors discussed the results and contributed to the manuscript.

**Funding Sources**

This work is partially funded by ERC advanced grant InSpectra.

**Notes**

The authors declare no competing financial interest.

## ACKNOWLEDGMENT

The authors acknowledge Andim Stassen, Wim Van Roy and Simone Severi in imec for fabrication of the waveguides, Rita Vos (imec), Tim Stakenborg (imec), Peter Bienstman (Photonics Research Group-UGent -imec) and Daan Martens (PRG-imec) for the functionalization of the chips with DNA and providing the cDNA. We also acknowledge Richard Penny (Center for nano and –biophotonics -UGent) for proofreading the manuscript, and Stéphane Clemmen (PRG-UGent-imec) and Ananth Subramanian (PRG-UGent-imec) for valuable discussions.

## ABBREVIATIONS

CCD: Charge Coupled Device

CMOS: Complementary Metal Oxide Semiconductor

IPA: Isopropyl alcohol

LFDC: Low Frequency Dominant Component

NWERS: Nanophotonic Waveguide Enhanced Raman spectroscopy

PECVD: Plasma Enhanced Chemical Vapor Deposition

SiN: Silicon nitride

SNR: Signal-to Noise Ratio

TE: quasi Transverse electric

TM: quasi Transverse Magnetic

## REFERENCES

1. Ulman, A., 2013. An Introduction to Ultrathin Organic Films: From Langmuir-Blodgett to Self-Assembly. Academic press.
2. Gennis, R.B. ed., 2013. Biomembranes: molecular structure and function. Springer Science & Business Media.
3. Richter, R.P., Bérat, R. and Brisson, A.R., 2006. Formation of solid-supported lipid bilayers: an integrated view. *Langmuir*, 22(8), pp.3497-3505.
4. Lu, B., Smyth, M.R. and O'Kennedy, R., 1996. Tutorial review. Oriented immobilization of antibodies and its applications in immunoassays and immunosensors. *Analyst*, 121(3), pp.29R-32R.
5. Heller, M.J., 2002. DNA microarray technology: devices, systems, and applications. *Annual review of biomedical engineering*, 4(1), pp.129-153.
6. Southern, E., Mir, K. and Shchepinov, M., 1999. Molecular interactions on microarrays. *Nature genetics*, 21, pp.5-9.
7. Castner, D.G. and Ratner, B.D., 2002. Biomedical surface science: Foundations to frontiers. *Surface Science*, 500(1), pp.28-60.
8. Chilkoti, A. and Hubbell, J.A., 2005. Biointerface science. *MRS Bulletin*, 30(03), pp.175-179.
9. Bonaccorso, F., Sun, Z., Hasan, T. and Ferrari, A.C., 2010. Graphene photonics and optoelectronics. *Nature photonics*, 4(9), pp.611-622.
10. Xu, M., Liang, T., Shi, M. and Chen, H., 2013. Graphene-like two-dimensional materials. *Chemical reviews*, 113(5), pp.3766-3798.
11. Castellanos-Gomez, A, 2016. Why all the fuss about 2D semiconductors? *Nature Photonics*, 10, 202–204
12. Poggi, M.A., Gadsby, E.D., Bottomley, L.A., King, W.P., Oroudjev, E. and Hansma, H., 2004. Scanning probe microscopy. *Analytical chemistry*, 76(12), pp.3429-3444.
13. Lee, P.A., Citrin, P.H., Eisenberger, P.T. and Kincaid, B.M., 1981. Extended x-ray absorption fine structure—its strengths and limitations as a structural tool. *Reviews of Modern Physics*, 53(4), p.769.
14. Cesareo, R., 2010. X-Ray Fluorescence Spectrometry. Wiley-VCH Verlag GmbH & Co. KGaA.
15. Ray, S., & Shard, A. G. Quantitative analysis of adsorbed proteins by X-ray photoelectron spectroscopy. *Analytical chemistry*, 83(22), 8659-8666 (2011)
16. Degen, C.L., Poggio, M., Mamin, H.J., Rettner, C.T. and Rugar, D., 2009. Nanoscale magnetic resonance imaging. *Proceedings of the National Academy of Sciences*, 106(5), pp.1313-1317.
17. Mrksich, M., Sigal, G.B. and Whitesides, G.M., 1995. Surface plasmon resonance permits in situ measurement of protein adsorption on self-assembled monolayers of alkanethiolates on gold. *Langmuir*, 11(11), pp.4383-4385.
18. Vollmer, F. and Arnold, S., 2008. Whispering-gallery-mode biosensing: label-free detection down to single molecules. *Nature methods*, 5(7), pp.591-596
19. Ferrari, A.C. and Basko, D.M., 2013. Raman spectroscopy as a versatile tool for studying the properties of graphene. *Nature nanotechnology*, 8(4), pp.235-246.
20. Kagan, M.R. and McCreery, R.L., 1995. Quantitative surface Raman spectroscopy of physisorbed monolayers on glassy carbon. *Langmuir*, 11(10), pp.4041-4047.
21. Kneipp, K., Kneipp, H., Itzkan, I., Dasari, R.R. and Feld, M.S., 2002. Surface-enhanced Raman scattering and biophysics. *Journal of Physics: Condensed Matter*, 14(18), p.R597.
22. Galarreta, B.C., Norton, P.R. and Lagugné-Labarthet, F., 2011. SERS Detection of Streptavidin/Biotin Monolayer Assemblies. *Langmuir*, 27(4), pp.1494-1498.
23. Lim, D.K., Jeon, K.S., Hwang, J.H., Kim, H., Kwon, S., Suh, Y.D. and Nam, J.M., 2011. Highly uniform and reproducible surface-enhanced Raman scattering from DNA-tailorable nanoparticles with 1-nm interior gap. *Nature nanotechnology*, 6(7), pp.452-460.
24. Liu, X., Shao, Y., Tang, Y. and Yao, K.F., 2014. Highly uniform and reproducible surface enhanced raman scattering on air-stable metallic glassy nanowire array. *Scientific reports*, 4.
25. Peyskens, F., Dhakal, A., Van Dorpe, P., Le Thomas, N. and Baets, R., 2016,. Surface enhanced Raman spectroscopy using a single mode nanophotonic-plasmonic platform. *ACS Photonics*. 3(1), pp 102-108
26. Opilik, L., Payamyar, P., Szczerbiński, J., Schütz, A.P., Servalli, M., Hungerland, T., Schlüter, A.D. and Zenobi, R., 2015. Minimally Invasive Characterization of Covalent Monolayer Sheets Using Tip-Enhanced Raman Spectroscopy. *ACS nano*, 9(4), pp.4252-4259.
27. Rabolt, J.F., Santo, R. and Swalen, J.D., 1980. Raman measurements on thin polymer films and organic monolayers. *Applied Spectroscopy*, 34(5), pp.517-521.
28. Kanger, J.S., Otto, C., Slotboom, M. and Greve, J., 1996. Waveguide Raman spectroscopy of thin polymer layers and monolayers of biomolecules using high refractive index waveguides. *The Journal of Physical Chemistry*, 100(8), pp.3288-3292.
29. Dhakal, A., Subramanian, A.Z., Wuytens, P., Peyskens, F., Le Thomas, N. and Baets, R., 2014. Evanescent excitation and collection of spontaneous Raman spectra using silicon nitride nanophotonic waveguides. *Optics letters*, 39(13), pp.4025-4028.
30. Dhakal, A., Raza, A., Peyskens, F., Subramanian, A.Z., Clemmen, S., Le Thomas, N. and Baets, R., 2015. Efficiency of evanescent excitation and collection of spontaneous Raman scattering near high index contrast channel waveguides. *Optics express*, 23(21), pp.27391-27404.
31. Dhakal, A., Peyskens, F., Clemmen, S., Raza, A., Wuytens, P., Zhao, H., Le Thomas, N. and Baets, R., 2016. Single mode waveguide platform for spontaneous and surface-enhanced on-chip Raman spectroscopy. *Interface Focus*, 6(4), p.20160015.
32. Evans, C.C., Liu, C. and Suntivich, J., 2016. TiO2 nanophotonic sensors for efficient integrated evanescent-Raman spectroscopy. *ACS Photonics*.
33. Holmstrom, S.A., Stievater, T.H., Kozak, D.A., Pruessner, M.W., Tyndall, N., Rabinovich, W.S., McGill, R.A. and Khurgin, J.B., 2016. Trace gas Raman spectroscopy using functionalized waveguides. *Optica*, 3(8), pp.891-896.
34. Subramanian, A.Z., Neutens, P., Dhakal, A., Jansen, R., Claes, T., Rottenberg, X., Peyskens, F., Selvaraja, S., Helin, P., Dubois, B. and Leyssens, K., 2013. Low-loss singlemode PECVD silicon nitride photonic wire waveguides for 532–900 nm wavelength window fabricated within a CMOS pilot line. *IEEE Photonics Journal*, 5(6), pp.2202809-2202809.
35. Jun, Y.C., Briggs, R.M., Atwater, H.A. and Brongersma, M.L., 2009. Broadband enhancement of light emission in silicon slot waveguides. *Optics Express*, 17(9), pp.7479-7490.
36. Ouldridge, T.E., Šulc, P., Romano, F., Doye, J.P. and Louis, A.A., 2013. DNA hybridization kinetics: zippering, internal displacement and sequence dependence. *Nucleic acids research*, 41(19), pp.8886-8895.

37. Kranenburg, O., 2005. The KRAS oncogene: past, present, and future. *Biochimica et Biophysica Acta (BBA)-Reviews on Cancer*, 1756(2), pp.81-82.
38. Okahata, Y., Kawase, M., Niikura, K., Ohtake, F., Furusawa, H. and Ebara, Y., 1998. Kinetic measurements of DNA hybridization on an oligonucleotide-immobilized 27-MHz quartz crystal microbalance. *Analytical Chemistry*, 70(7), pp.1288-1296.
39. Jeng, E.S., Barone, P.W., Nelson, J.D. and Strano, M.S., 2007. Hybridization Kinetics and Thermodynamics of DNA Adsorbed to Individually Dispersed Single-Walled Carbon Nanotubes. *Small*, 3(9), pp.1602-1609.
40. Sorgenfrei, S., Chiu, C.Y., Gonzalez Jr, R.L., Yu, Y.J., Kim, P., Nuckolls, C. and Shepard, K.L., 2011. Label-free single-molecule detection of DNA-hybridization kinetics with a carbon nanotube field-effect transistor. *Nature nanotechnology*, 6(2), pp.126-132.
41. Hermanson, G.T., 2013. Bioconjugate techniques. Academic press.
42. Weber, P.C., Ohlendorf, D.H., Wendoloski, J.J. and Salemme, F.R., 1989. Structural origins of high-affinity biotin binding to streptavidin. *Science*, 243(4887), pp.85-88.
43. Diamandis, E.P. and Christopoulos, T.K., 1991. The biotin-(strept) avidin system: principles and applications in biotechnology. *Clinical chemistry,* 37(5), pp.625-636.
44. Langer, P.R., Waldrop, A.A. and Ward, D.C., 1981. Enzymatic synthesis of biotin-labeled polynucleotides: novel nucleic acid affinity probes. *Proceedings of the National Academy of Sciences*, 78(11), pp.6633-6637.
45. Bistričić, L., Volovšek, V. and Dananić, V., 2007. Conformational and vibrational analysis of gamma-aminopropyltriethoxysilane. *Journal of molecular structure*, 834, pp.355-363.
46. Fagnano, C., Fini, G. and Torreggiani, A., 1995. Raman spectroscopic study of the avidin—biotin complex. *Journal of Raman Spectroscopy*, 26(11), pp.991-995.
47. Shim, S., Stuart, C.M. and Mathies, R.A., 2008. Resonance Raman Cross-Sections and Vibronic Analysis of Rhodamine 6G from Broadband Stimulated Raman Spectroscopy. *ChemPhysChem*, 9(5), pp.697-699.
48. Kagan, M.R. and McCreery, R.L., 1994. Reduction of fluorescence interference in Raman spectroscopy via analyte adsorption on graphitic carbon. Analytical Chemistry, 66(23), pp.4159-4165.
49. Hoff, J.D., Cheng, L.J., Meyhöfer, E., Guo, L.J. and Hunt, A.J., 2004. Nanoscale protein patterning by imprint lithography. *Nano letters*, 4(5), pp.853-857.
50. Arafat, A., Giesbers, M., Rosso, M., Sudhölter, E.J., Schroën, K., White, R.G., Yang, L., Linford, M.R. and Zuilhof, H., 2007. Covalent biofunctionalization of silicon nitride surfaces. *Langmuir*, 23(11), pp.6233-6244.
51. Zhang, F. and Srinivasan, M.P., 2004. Self-assembled molecular films of aminosilanes and their immobilization capacities. *Langmuir*, 20(6), pp.2309-2314.

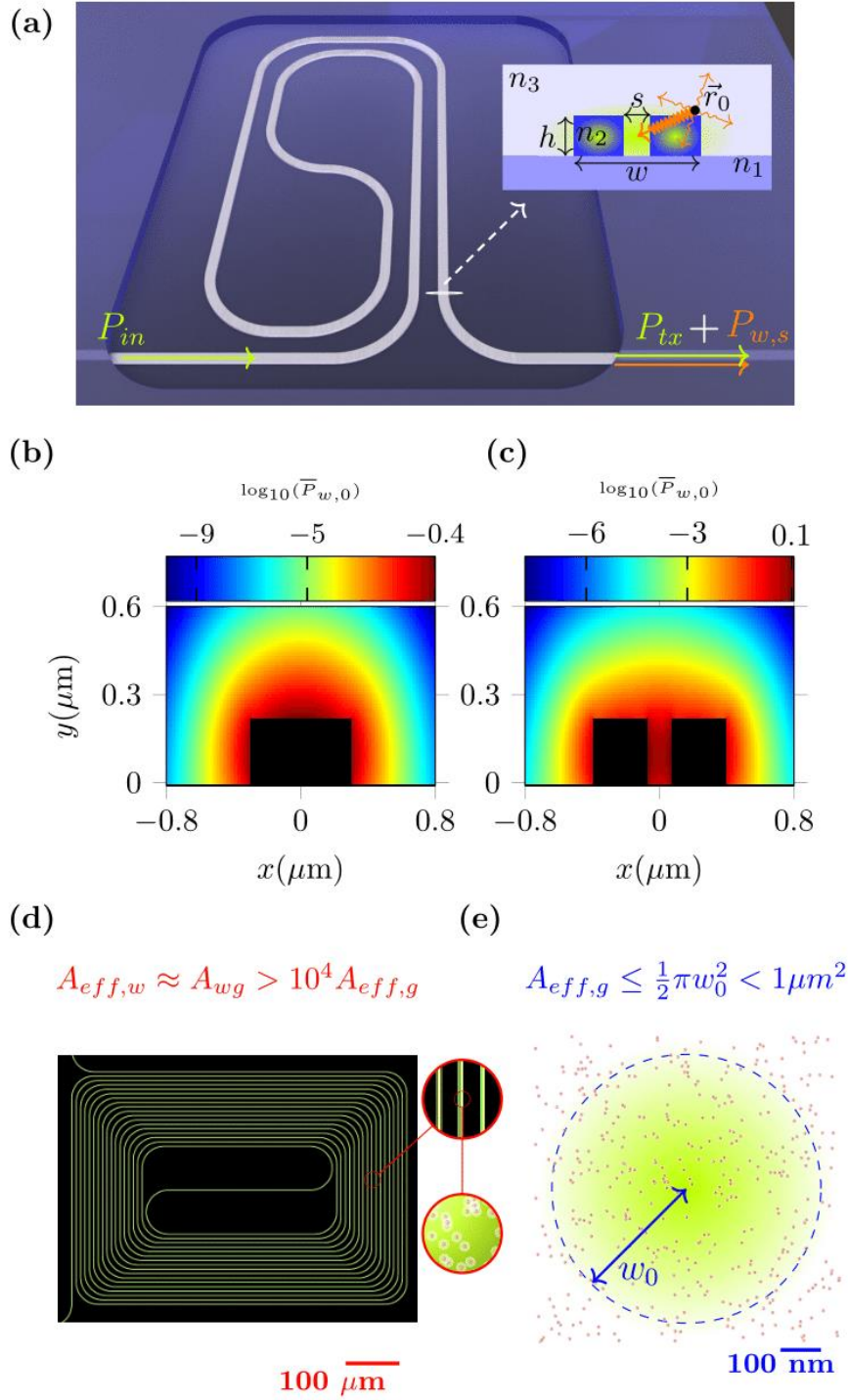

**Figure 1| Schematic and principle of NWERS. (a) The schematic of the NWERS system.** Top inset shows a generic slot waveguide. For the waveguide used in this article $n_1$=1.45 (SiO$_2$), $n_2$=1.89 (Si$_3$N$_4$), $n_3$=1.33 (H$_2$O), $h$ = 220 nm. The map of $log_{10}(\overline{P}_{w,o})$, for different positions of a molecule in the top cladding region for (b) strip ($w$ = 600 nm, $s$ =0 nm) and (c) slot waveguides ($w$ = 850 nm, $s$ =150 nm) investigated in this article. $\overline{P}_{w,o}$ is the power coupled from a particle to the fundamental TE waveguide mode, normalized to the total emission from the particle when excited by a diffraction limited beam with NA=1. (d) Illustration of a typical 1 cm silicon nitride waveguide spiral used in NWERS. The effective interaction area, $A_{eff,w}$ is nearly equal to the physical area $A_{wg}$ of the waveguide which is more than four orders of magnitude larger than the effective area $A_{eff,g}$ of a diffraction limited beam shown in (e).

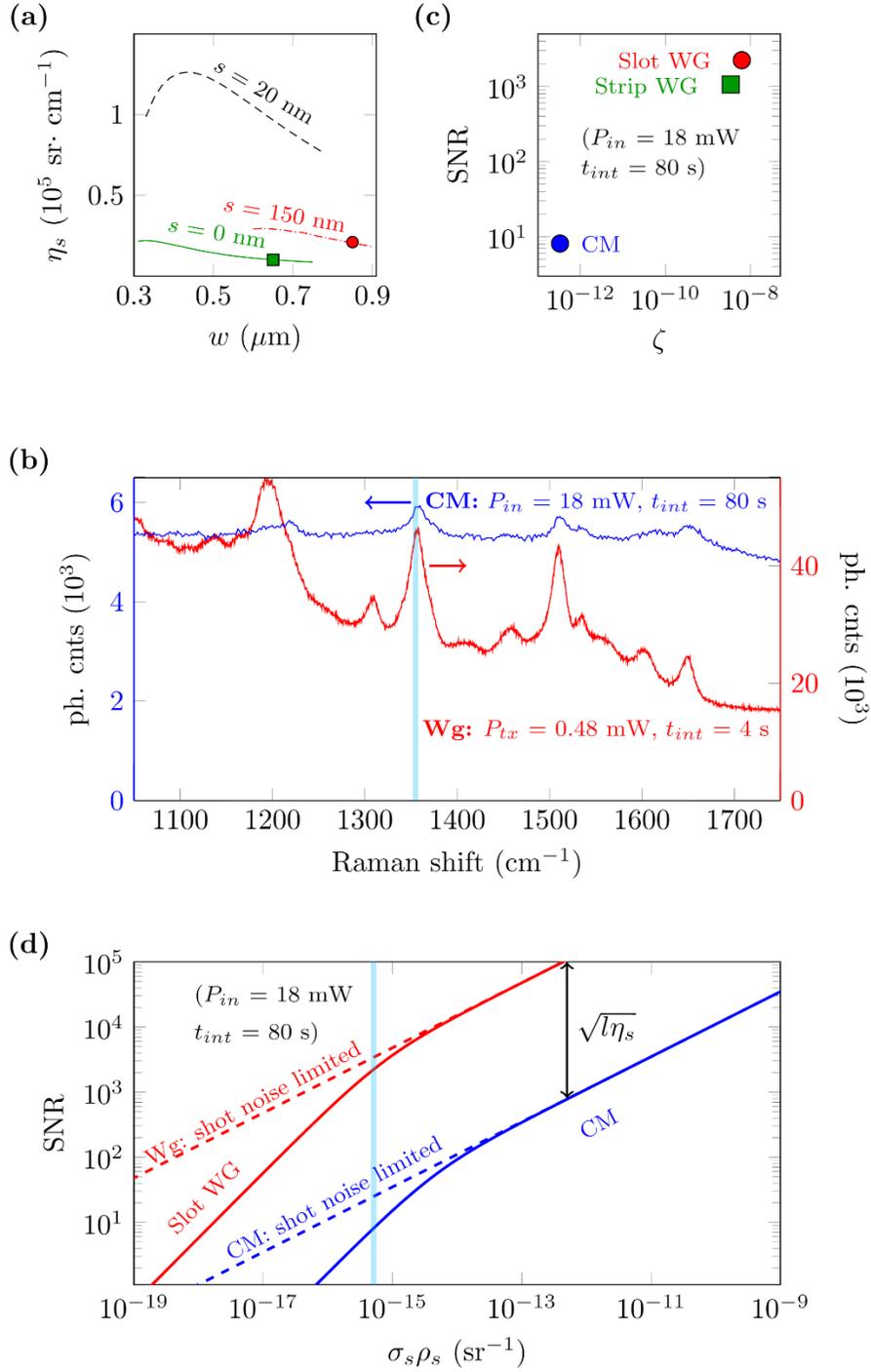

**Figure 2| Enhancement of signal and improvement of SNR with NWERS.** (a) The calculated values of the $\eta_s$ for NWERS system for $h$ = 220 nm, slot widths $s$ = 0, 20 nm and 150 nm as a function of waveguide width $w$. The waveguides used in the experiments are marked with circles. (b) The measured spectra of a Rhodamine monolayer obtained from a commercial Raman microscope (CM, $P_{in}$ = 18 mW and $t_{int}$ = 80 s, in blue with left blue axis) in contrast to the spectra obtained from the 1 cm slot waveguides (Wg, $P_{in}$= 30 mW, $P_{tx}$= 0.48 mW and $t_{int}$= 4 s, in red with right red axis). (c) SNR vs conversion ratio $\zeta$ for the 1355 cm$^{-1}$ line (highlighted with cyan in (b)) from rhodamine monolayers obtained from a commercial microscope (CM), slotted waveguides (red) and strip waveguides (green). (d) Evolution of SNR for different $\rho_s\sigma_s$ based on the data presented in (b) and highlighted by a cyan bar in the figure (solid lines). Evolution of SNR for the ideal cases, when shot-noise from the signal is the only source of noise, is also shown in dashed lines for the respective systems

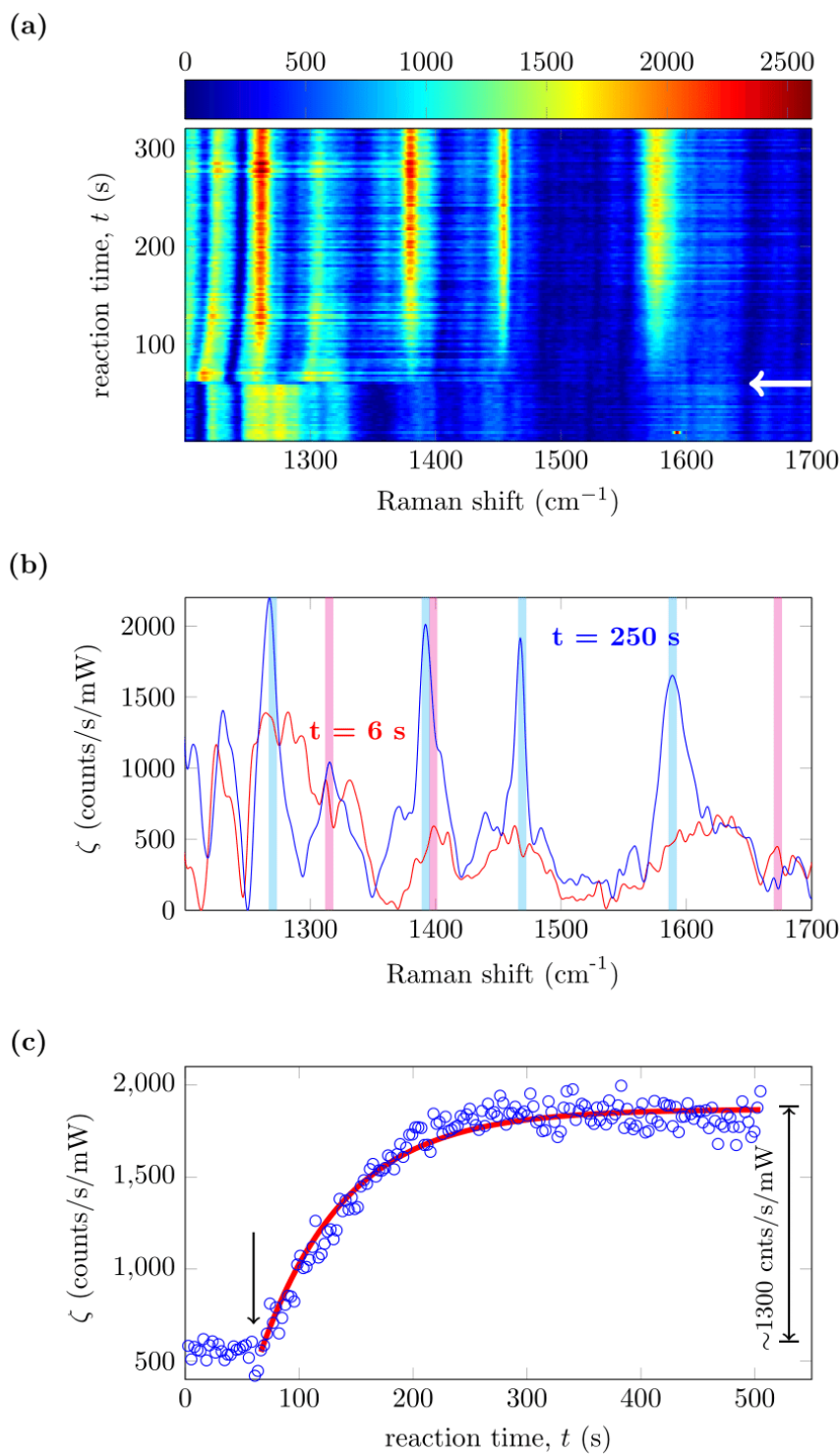

**Figure 3| Real time observation of DNA hybridization process using NWERS.** (a) Raman spectrogram as a function of reaction time $t$. cDNA-Cy3 was added at $t = 60$ s (indicated by an arrow). (b) The Raman spectra before addition of the cDNA ($t = 6$ s, red) and after the saturation of hybridization reaction ($t = 250$ s, blue). Cyan and magenta are respectively Raman lines from cy3 and nucleic acids (c) The Raman signal (blue circles), corresponding to the 1392 cm$^{-1}$ line of Cy3, follows the first order reaction equation (red line) with goodness of fit $R^2 = 0.952$.

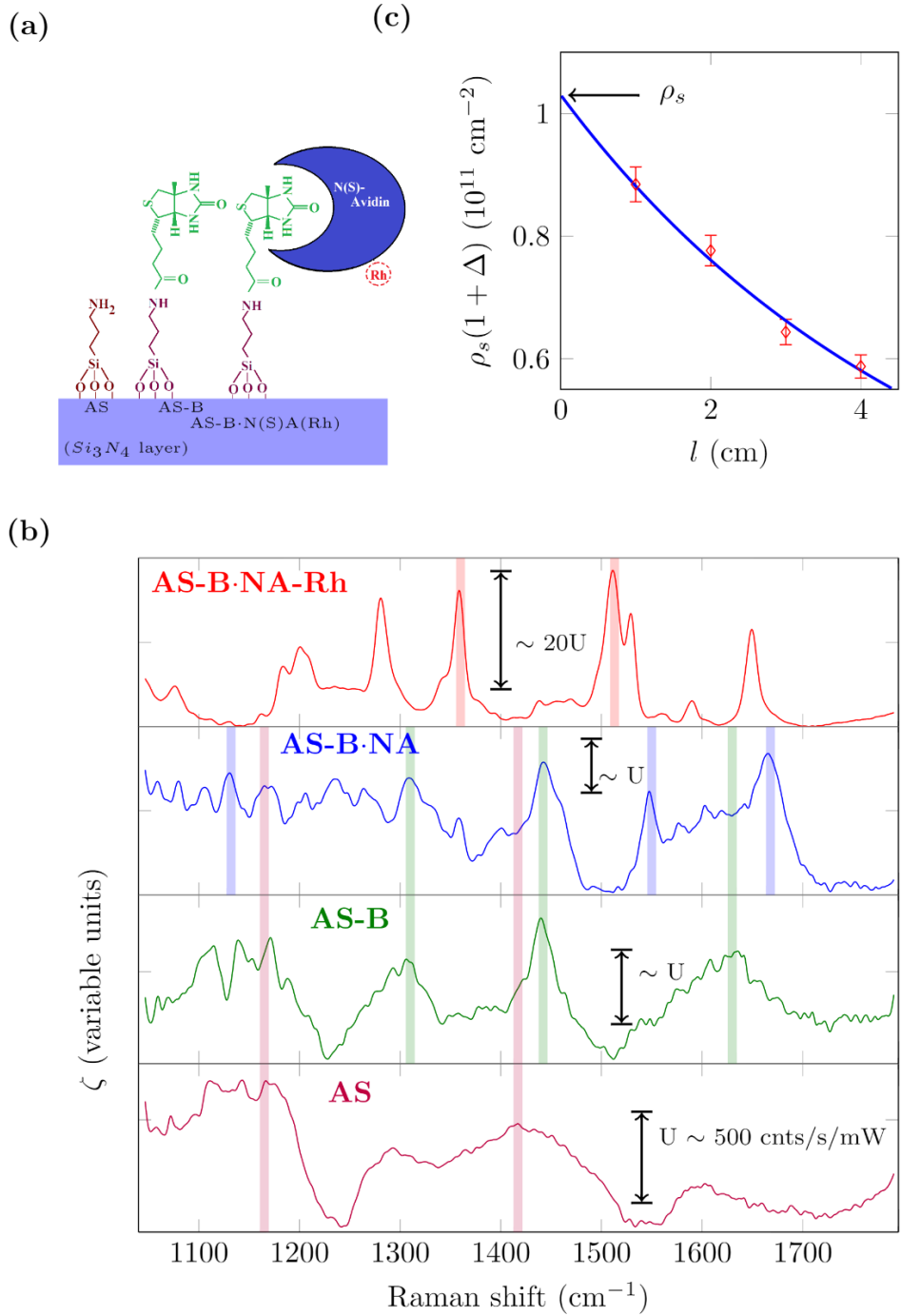

**Figure 4| Detection and quantification of Biotin-Avidin monolayers** (a) Schematic of immobilized monolayers at different stages of functionalization. APTES amino-silane (AS), biotin (B) covalently bound to AS (AS-B), NeutrAvidin (NA) or Streptavidin conjugated with Rhodamine (SA-Rh) attached to AS-B with non-covalent bond (AS-B· NA(SA-Rh)) are depicted. (b) Raman spectrum of AS (brown), AS-B (green), AS-B· NA (blue) and B-AS-B·SA-Rh (red). The vertical transparent lines of respective colors indicate the Raman peaks due to the corresponding molecules as found in the literature. (c) Dependence of the $\rho_s(1+\Delta)$ as a function of waveguide length calculated using 1513 cm$^{-1}$ Rhodamine Raman signal obtained from B-AS-B·SA-Rh attached to biotin. The red diamonds are the experimental values and the blue line is the least squared error fit with Eq. (2) as a model. Estimated molecular density $\rho_s$ is shown by an arrow corresponding to $l \sim 0$, where the effect of $\Delta$ is negligible.

# Nanophotonic waveguide enhanced Raman spectroscopy of biological submonolayers

Ashim Dhakal[1, 2, 3,*], Pieter C. Wuytens[1,2,4], Frédéric Peyskens[1,2], Karolien Jans[3], Nicolas Le Thomas[1,2] and Roel Baets[1,2]

## Supplementary Information

### 1. Normalized power scattered by a particle and surface conversion efficiency

Here we outline a mathematical model to calculate the scattered power coupled to a mode of an arbitrary dielectric channel waveguide and compare it with the case with diffraction-limited beams in the free-space.

We consider a molecule of scattering cross-section $\sigma$ located at an arbitrary location $r_0$ near a waveguide with its fundamental mode carrying a guided pump of power $P_{in}$. The total power $P_{w,0}(r_0)$ of the scattered light coupled to the same waveguide mode is proportional to the fourth power of the power-normalized modal $e_m(r_0)$ field and given by Eq. (SE1)[1].

$$P_{w,0}(\vec{r}_0) = \Lambda(\vec{r}_0)\sigma P_{in} \tag{SE1}$$

Specifically, using a semi-classical perturbative approach[1], $\Lambda$ can be approximated by:

$$\Lambda(\vec{r}_0) = \frac{\lambda_0^2 n_g^2}{n(\vec{r}_0)} \left( \frac{|\vec{e}_m(\vec{r}_0)|^2}{\iint_\infty n(\vec{r})^2 |\vec{e}_m(\vec{r})|^2 d\vec{r}} \right)^2 \tag{SE2}$$

where, $n_g$ is the group index of the mode, $\lambda_0$ is the wavelength of the pump and the Stokes light assumed to be sufficiently close to the pump (for the waveguides and wavelengths used in this paper, the assumption that emission wavelength $\sim \lambda_0$ leads to an average underestimation $< 5\%$ for emissions within Raman shifts $< 1500$ cm$^{-1}$), $n(\mathbf{r})$ is the refractive index function.

The total Raman power emitted by the same molecule when excited by the average field intensity within the beam waist ($w_0$) of diffraction-limited beam of unit numerical aperture (the most ideal situation for a free-space beam) carrying the same pump power is given by:

$$P_{g,0} = \frac{P_{in}}{\pi w_0^2}\sigma = \frac{\pi P_{in}}{\lambda_0^2}\sigma \tag{SE3}$$

We choose a diffraction-limited beam as the ideal beam for Raman microscopes, since it has the minimum possible étendue. Minimum étendue ensures maximal power density for maximal excitation. To compare with the ideal free-space excitation for Raman microscope, $P_{w0}(\mathbf{r}_0)$ can be normalized with $P_{g,0}$:

$$\bar{P}_{w,0}(\vec{r}_0) = \frac{P_{w,0}(\vec{r}_0)}{P_{g,0}} = \frac{\lambda_0^2}{\pi}\Lambda(\vec{r}_0) \tag{SE4}$$

Figure (SF1) plots $log_{10}(\bar{P}_{w,0})$ for different positions in the surrounding of a silicon nitride slot waveguide ($s = 20$ nm, $w = 660$ nm) in water calculated by the *COMSOL* finite elements mode solver for the fundamental TE mode. The figure shows that more than one order of magnitude of power is coupled to the waveguide mode compared to the total emission in the most ideal diffraction-limited system. This result demonstrates the possibility of a large broadband Purcell enhancement[2] that could be utilized in NWERS.

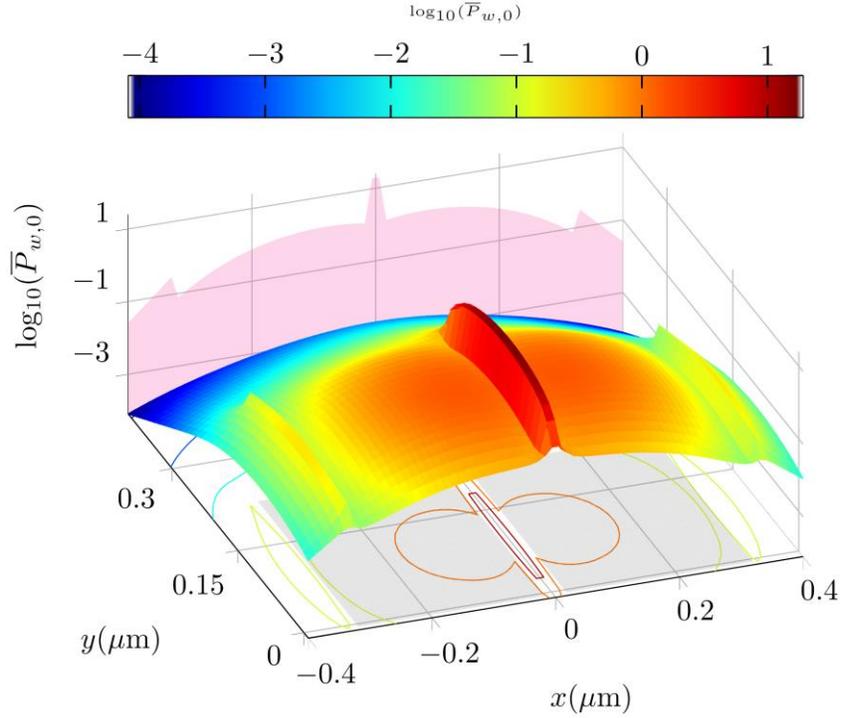

**Figure SF1| Normalized power coupled to the waveguide mode**. $log_{10}(\bar{P}_{w,0})$ for a slotted waveguide ($s$ = 20 nm, $w$ = 660 nm) shown in 3D highlighting the variations of the coupled power across the various regions of the section. $\bar{P}_{w,0}$ is the power coupled from a particle to the fundamental TE waveguide mode, normalized to the total emission from the particle when excited by a diffraction limited beam with NA=1. The purple shaded area shows the projection of $\bar{P}_{w,0}(\mathbf{r}_0)$, indicating the maxima of the $\bar{P}_{w,0}(\mathbf{r}_0)$. The gray shaded region in the x-y plane indicates the waveguide cross-section. The contours represent the lines with the same $\bar{P}_{w,0}$ and the color bar shows the magnitude of $\bar{P}_{w,0}$.

For a monolayer of molecules with surface density $\rho_s$ uniformly distributed over a transversely symmetric waveguide of length $l$, the total normalized surface conversion efficiency for the waveguide system is given by:

$$\frac{P_{w,s}}{P_{in}} = \sigma \rho_s l \int_{line} \Lambda(\vec{r}) d\vec{r} \qquad (SE5)$$
$$= l \rho_s \sigma \eta_s$$

The line integral in Eq. (SE5) is defined along the interface between the core and upper cladding on an arbitrary section of a waveguide (Fig. SF2). We call the integral, $\eta_s$ the *surface conversion efficiency,* which is given by Eq. (SE6).

$$\eta_s \equiv \int_{line} \Lambda(\vec{r}_0) d\vec{r} \qquad (SE6)$$

For a similar monolayer excited and collected by a diffraction limited free-space system, the interaction area is limited by the beam waist. Integrating Eq. (SE3), the power emitted *in every direction* when excited by a diffraction-limited beam is given by $P_{g,s} = P_{pump} \sigma \rho_s$.

Hence,

$$\bar{P}_{w,s} \equiv \frac{P_{w,s}}{P_{g,s}} = l \eta_s \qquad (SE7)$$

Fig. 2(a) in the main article show the $\eta_s$ calculated for silicon nitride strip and slot waveguides using the COMSOL mode solver for various geometries.

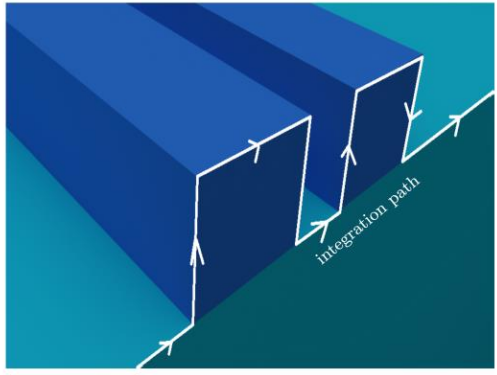

**Figure SF2│ Integration path to calculate $\eta_s$.** The surface integral can be reduced to a line integral along the path (shown in white) for uniformly distributed molecules on top of the longitudinally invariant waveguide.

## 2. Considerations for SNR.

In this section, we discuss the signal-to-noise performance of the background shot-noise limited NWERS system for low concentration in comparison with the dark-noise limited free-space system.

We verified that the photon counts statistics of the CCD detector follows a Poisson distribution. It follows that the SNR can be defined as:

$$SNR \equiv \frac{C_\nu - C_{BG}}{\sqrt{C_\nu}} \qquad (SE8)$$

where, $C_\nu$ is the average number of photon counts of the signal peak at wavenumber $\nu$ and $C_{BG}$ is the average number of photon counts near the bottom of the peak that correspond to the background (see section 5 for further details). In terms of experimental variables such as input power $P_{in}$, signal integration time $t_{int}$, the product of scattering cross-section and the density of the molecules ($\beta_s$ for signal and $\beta_{BG}$ for the back ground), interaction length $l$ and detector dark noise equivalent power $D$, the SNR can be expressed as:

$$SNR = \theta \frac{P_{in} \eta_s l \beta_s \sqrt{t_{int}}}{\sqrt{P_{in} l (\eta_s \beta_s + \eta_{BG} \beta_{BG}) + D}} \qquad (SE9)$$

Where, $\theta$ is square root of the detector sensitivity which will be omitted in the following. Here, we have considered the waveguide materials act as a source of background light emitting in the same frequency, which is quantified by $\eta_{BG}\beta_{BG}$.

**High concentration limit**

If the shot noise originating from the signal is the dominant source of noise, i.e. $\beta_s \eta_s > \beta_{BG} \eta_{BG} > D$, Eq. (SE9) can be written as:

$$SNR = \sqrt{l P_{in} t_{int} \eta_s \beta_s} \qquad (SE10)$$

In this situation, the SNR for the NWERS system is larger than the ideal free-space system by a factor $\sqrt{(l\eta_s)}$ for the same integration time and input power. To obtain a similar SNR, the integration time needed is then reduced by a factor $(l\eta_s)$ for the NWERS system compared to an ideal free-space system.

**Low concentration limit**

For a waveguide system detecting a very low analyte concentration, $\beta_{BG}$ is the dominating source of noise, hence, the SNR for the waveguide $SNR_{WG}$ can be written as:

$$SNR_{WG} = \sqrt{l t_{int} P_{in}} \frac{\eta_s \beta_s}{\sqrt{\eta_{BG} \beta_{BG}}} \qquad (SE11)$$

Similarly for an ideal free-space system without any source of background, SNR is limited only by the dark noise of the detector; hence the SNR for a dark limited system $SNR_{DL}$ is given by

$$SNR_{DL} = P_{in} \beta_\nu \sqrt{\frac{t_{int}}{D}} \qquad (SE12)$$

Hence, for the same integration time and input power,

$$\frac{SNR_{WG}}{SNR_{DL}} = \eta_s \sqrt{\frac{l}{P_{in}} \frac{D}{\eta_{BG}\beta_{BG}}}$$ (SE13)

For a low concentration, and similar SNR, the reduction in integration time for a NWERS system compared to the ideal free-space system is a quadratic function of the surface conversion efficiency $\eta_s$ since

$$\frac{t_{int,DL}}{t_{int,WG}} = \eta_s^2 \frac{l}{P_{in}} \frac{D}{\eta_{BG}\beta_{BG}}$$ (SE14)

From our measurements, we have estimated $\eta_{BG}\beta_{BG} \sim 10^{-9}$ cm$^{-1}$ and $D \sim 2$ fW for the CCD used in our measurement operating at maximum cooling at -80° C. Then, for 1cm waveguide and 1 mW pump, the reduction in integration time is $t_{DL}/t_{WG} \sim 10^{-3} \eta_s^2$. For the waveguide investigated in this paper, $\eta_s \sim 2 \cdot 10^5$. Hence, the reduction in the integration time is $\sim 8 \cdot 10^5$.

## 3. DNA Hybridization kinetics.

Here we develop a model of Raman signal dependence as a function of DNA hybridization. We consider a probe monolayer of DNA immobilized on top of the photonic chip with a total surface density $\rho_T$. The free (unoccupied) immobilized probe DNA with average surface density $\rho_F$ and the target cDNA in the analyte solution with surface density $\rho_{A,s}$ (with a proportional volume density $\rho_A$) produce the hybridization duplex with surface density $\rho_H$, such that $\rho_T = \rho_H + \rho_F$. The reaction can be described by the following kinetic equation.

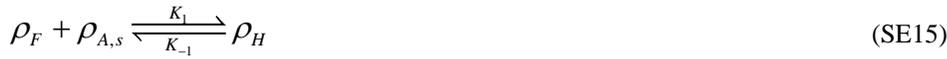
(SE15)

Here $K_1$ is the association constant that describes the binding rate of DNA and cDNA, while $K_{-1}$ is the dissociation constant which describes the rate of dissociation of hybridization duplex into the surface DNA and cDNA. Assuming no other interactions between the species occur, the hybridization rate can be described by a first order rate equation as follows.

$$\frac{d\rho_H}{dt} = K_1 \rho_F \rho_A - K_{-1} \rho_H$$ (SE16)

The unoccupied DNA density $\rho_F$ participating in the reaction is the difference between total probe DNA concentration $\rho_T$ and hybridized DNAs concentration, i.e $\rho_F = \rho_T - \rho_H$. Thus,

$$\frac{d\rho_H}{dt} = -\rho_H \left( K_1 \rho_A + K_{-1} \right) + \rho_A \rho_T K_1$$ (SE17)

We assume that the concentration $\rho_A$ of the target cDNA on top of the waveguide remains constant and is sufficiently low to affect the signal from hybridization complex. Then, the measured Raman signal from the hybridization is proportional to the density of the hybridization complex $\rho_H$ as per Eq. (2) of the main article. Hence, we obtain Eq. (SE18) for the evolution of the Raman signal.

$$\zeta(\tau) = \zeta_\infty \left( 1 - e^{-(K_1 \rho_A + K_{-1})\tau} \right)$$ (SE18)

where, $\zeta(\tau)$ is the normalized Raman signal intensity, defined by Eq. (2) in the main article, collected through the waveguide at a time $\tau$ after the start of the reaction, $\zeta_\infty$ is the normalized Raman signal corresponding to the total number of binding sites i.e. the Raman signal at the time of saturation.

For the hybridization of DNA with a chain length > 20, as in our case, typically[3,4] $K_{-1} < 10^{-4}$ s$^{-1}$ while $K_1 > 10^5$ M$^{-1}$s$^{-1}$, thus, for a concentration $\rho_A > 10$ nM, the equation is reduced to:

$$\zeta(\tau) = \zeta_\infty \left( 1 - e^{-\tau K_1 \rho_A} \right)$$ (SE19)

## 4. Determination of the cross-section of cy3

The value of $\zeta_\infty$ corresponding to the 1392 cm$^{-1}$ line of Cy3 molecules was experimentally determined to be $1.3 \cdot 10^3$ counts/s/mW (Fig 3(c)). The detector sensitivity is determined to be $1.9 \cdot 10^{15}$ counts/s/mW. This allows us to calculate $\zeta_\infty = 6.7 \cdot 10^{-13}$ in absolute units. We take $\eta_s = 2.1 \cdot 10^4$ sr·cm$^{-1}$ obtained from the simulations. As discussed in the main text, the surface efficiency value $\eta_s$ is tolerant to small variations in the waveguide that might occur during fabrication hence we expect $\eta_s$ to be very close to the simulated value[1]. X-ray fluorescence spectroscopy data provides us the density of the DNA probes $\rho_s = 6 \pm 1 \cdot 10^{-13}$. Hence, using Eq. (2) for a 1 cm long waveguide, we determine $\sigma(1+\Delta)$ for the 1392 cm$^{-1}$ line of Cy3 to be $1.1 \pm 0.2 \cdot 10^{-29}$ cm$^2$·sr$^{-1}$·molecule$^{-1}$ As seen Fig. 4 (c)), $\Delta$ accounts for about 15% underestimation of the parameters for the 1 cm waveguide. Thus, $\sigma = 1.3 \pm 0.2 \cdot 10^{-29}$ cm$^2$·sr$^{-1}$·molecule$^{-1}$ for the 1392 cm$^{-1}$ line of Cy3 when pumped at 785 nm.

## 5. Background modeling and subtraction algorithm.

In sec. 2 of this supplementary information, we saw that the SNR performance of the NWERS system is superior compared to an idealized free-space system despite the background from the waveguide. Nevertheless, any existing background due to the waveguide materials needs to be subtracted, particularly in the case of low concentration when the signal from the analyte is weak. Fortunately, the waveguide background is generally constant, can be well-characterized and modelled. In this section we describe a simple model of the background and a simple background subtraction algorithm used throughout the article. We model the measured NWERS signal $S_{MES}$ as:

$$S_{MES}(\lambda) = R_{ANL}(\lambda) + P_{SIN+ANL}(\lambda) + R_{SIN}(\lambda) \tag{SE20}$$

Here $R_{ANL}$ is the Raman signal exclusively from the analyte, $P_{SIN+ANL}$ is the polynomial background usually consisting of auto-fluorescence from the $Si_3N_4$ core and the analyte, $R_{SIN}$ is the Raman signal from the $Si_3N_4$ core[5]. A reference spectrum from the waveguide, measured without the analyte is:

$$S_{REF}(\lambda) = P_{SIN}(\lambda) + \alpha R_{SIN}(\lambda) \tag{SE21}$$

The polynomial backgrounds $P_{SIN}$ and $P_{SIN+ANL}$ can be approximated by using the asymmetric cost function algorithm[6] (in our data we use 3$^{rd}$ order polynomials). After subtraction of the background polynomials this becomes:

$$\tilde{S}_{MES} = S_{MES} - P_{SIN+ANL} = R_{ANL} + R_{SIN} + e_1 \tag{SE22}$$

and,

$$\tilde{S}_{REF} = S_{REF} - P_{SIN} = \alpha R_{SIN} + e_2 \tag{SE23}$$

where $e_1$ and $e_2$ are small residual errors that might have remained during the subtraction process. The Raman signal contribution from the core $R_{SIN}$ in Eq. (SE21) can now be obtained from the $\tilde{S}_{REF}$ by calculating the linear scaling factor $\alpha$ using least square algorithm that minimizes the cost defined as follows:

$$C = \left( \tilde{S}_{MES}(\lambda) - \frac{\tilde{S}_{REF}(\lambda)}{\alpha} \right)^2 \tag{SE24}$$

An approximate Raman spectrum of the analyte on top of the waveguide is then given by:

$$R_{ANL} = \tilde{S}_{MES} - \frac{\tilde{S}_{REF}}{\alpha} + e \tag{SE25}$$

The approximate spectrum may contain some residual background $e$ that has not been completely removed during the process. The residual background $e$ is then removed using a variant of the low order asymmetric Whittaker method developed by Eilers[7].

Figure (SF4) illustrates the background subtraction algorithm just discussed. First the reference spectrum $S_{REF}$ of the waveguides without the analyte is measured as shown in Fig. SF4(a). The third order polynomial part of the background, obtained using the asymmetric cost function algorithm is then subtracted yielding an approximate background $\tilde{S}_{REF}$ (Fig. SF4(b)). The polynomial-subtracted background consists mainly of a homogeneously broadened Raman spectrum $R_{SIN}$ of the deposited $Si_3N_4$. This Raman spectrum of the waveguide is then rescaled, as necessary, to the similar spectra $\tilde{S}_{MES}$ obtained with analyte and subtracted yielding an approximate signal spectrum of the analyte $R_{ANL}$. The residual $e$ that may have cropped in during the process is removed to get the final spectra as shown in Fig SF4 (d).

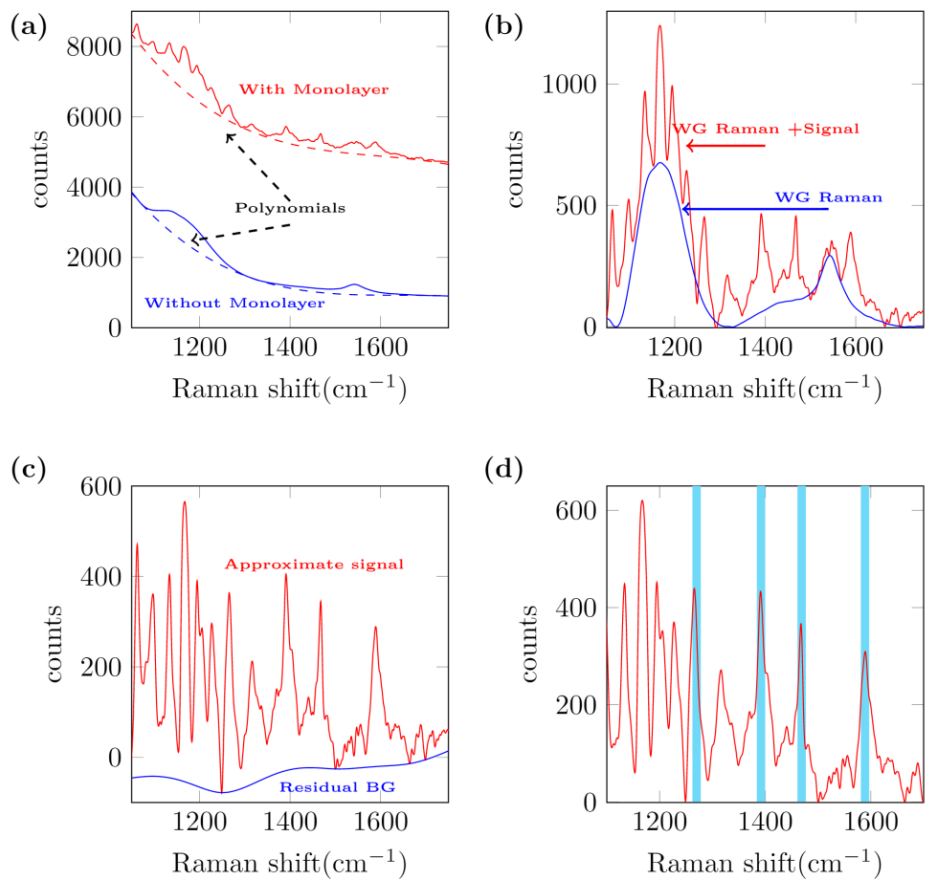

**Figure SF4| Illustration of the background subtraction algorithm as applied to a monolayer of DNA·cDNA-Cy3 hybridization.** (**a**) Raw spectrum of the waveguides with and without monolayers and the corresponding third order polynomial approximations of the polynomial component of the background. (**b**) The difference of the spectra and corresponding polynomials shown in (a). The difference spectrum obtained from the waveguide without analyte is rescaled to that with analyte. (**c**) The difference between the rescaled background without and with the analyte, giving an approximate Raman spectrum of the analyte alone. Any residual background is subtracted to obtain the final spectrum of the analyte shown in (d). **d**) The final Raman spectrum indicating the major Raman lines corresponding to Cy3.

## References

1. Dhakal, A., Raza, A., Peyskens, F., Subramanian, A.Z., Clemmen, S., Le Thomas, N. and Baets, R., 2015. Efficiency of evanescent excitation and collection of spontaneous Raman scattering near high index contrast channel waveguides. Optics express, 23(21), pp.27391-27404.
2. Jun, Y.C., Briggs, R.M., Atwater, H.A. and Brongersma, M.L., 2009. Broadband enhancement of light emission in silicon slot waveguides. Optics Express, 17(9), pp.7479-7490.
3. Okahata, Y., Kawase, M., Niikura, K., Ohtake, F., Furusawa, H. and Ebara, Y., 1998. Kinetic measurements of DNA hybridization on an oligonucleotide-immobilized 27-MHz quartz crystal microbalance. Analytical Chemistry, 70(7), pp.1288-1296.
4. Jeng, E.S., Barone, P.W., Nelson, J.D. and Strano, M.S., 2007. Hybridization Kinetics and Thermodynamics of DNA Adsorbed to Individually Dispersed Single-Walled Carbon Nanotubes. Small, 3(9), pp.1602-1609
5. Smith, D.L., Alimonda, A.S., Chen, C.C., Ready, S.E. and Wacker, B., 1990. Mechanism of SiN x H y Deposition from NH 3-SiH4 Plasma. Journal of the Electrochemical Society, 137(2), pp.614-623.
6. Mazet, V., Carteret, C., Brie, D., Idier, J. and Humbert, B., 2005. Background removal from spectra by designing and minimising a non-quadratic cost function. Chemometrics and intelligent laboratory systems, 76(2), pp.121-133.
7. Eilers, P.H. and Boelens, H.F., 2005. Baseline correction with asymmetric least squares smoothing. Leiden University Medical Centre Report.